\documentclass[aps,prl,twocolumn,superscriptaddress]{revtex4-1}
\usepackage{graphicx,subfigure}
\usepackage{amsmath,amssymb}
\usepackage{mathtools}
\usepackage{multirow}
\usepackage{textcomp}
\usepackage{float}
\usepackage{color}
\usepackage{dsfont}
\usepackage{tikz}
\usepackage{enumitem}
\usetikzlibrary{math}
\setlist{noitemsep}
\usepackage[colorlinks,linkcolor=blue,citecolor=blue,urlcolor=blue,breaklinks=true]{hyperref}

\newcommand{\avg}[1]{\langle #1 \rangle}
\newcommand{\bra}[1]{\ensuremath{{\left\langle #1 \right|}}}
\newcommand{\ket}[1]{\ensuremath{{\left| #1 \right\rangle}}}
\renewcommand{\vec}[1]{\ensuremath{\mathbf{#1}}}
\newcommand{\uvec}[1]{\ensuremath{\mathbf{\hat{#1}}}}
\newcommand{\abs}[1]{\ensuremath{\vert #1 \vert}}

\newcommand{\C}{\ensuremath{C_\mathrm{cl}}}

\newcommand{\dgraph}{\ensuremath{r_{ij}}}
\newcommand{\vs}{vs. }
\newcommand{\GscC}{\ensuremath{\Gamma_\mathrm{sc,0}}}
\newcommand{\Gsc}{\ensuremath{\Gamma_\mathrm{sc}}}
\newcommand{\Op}{\mathcal{O}}
\newcommand{\adj}[1]{#1^{\dagger}}

\graphicspath{{./}}

\begin{document}

\title{Treelike Interactions and Fast Scrambling with Cold Atoms}

\author{Gregory Bentsen}
\affiliation{Department of Physics, Stanford University, Stanford, CA 94305, USA}

\author{Tomohiro Hashizume}
\author{Anton S.~Buyskikh}
\affiliation{Department of Physics and SUPA, University of Strathclyde, Glasgow G4 0NG, UK}

\author{Emily J.~Davis}
\affiliation{Department of Physics, Stanford University, Stanford, CA 94305, USA}

\author{Andrew J.~Daley}
\affiliation{Department of Physics and SUPA, University of Strathclyde, Glasgow G4 0NG, UK}

\author{Steven S.~Gubser}
\affiliation{Department of Physics, Princeton University, Princeton, NJ 08544, USA}

\author{Monika Schleier-Smith}
\affiliation{Department of Physics, Stanford University, Stanford, CA 94305, USA}

\date{\today}

\begin{abstract}
We propose an experimentally realizable quantum spin model that exhibits fast scrambling, based on non-local interactions which couple sites whose separation is a power of $2$.  By controlling the relative strengths of deterministic, non-random couplings, we can continuously tune from the linear geometry of a nearest-neighbor spin chain to an ultrametric geometry in which the effective distance between spins is governed by their positions on a tree graph.  The transition in geometry can be observed in quench dynamics, and is furthermore manifest in calculations of the entanglement entropy.  Between the linear and treelike regimes, we find a peak in entanglement and exponentially fast spreading of quantum information across the system.  Our proposed implementation, harnessing photon-mediated interactions among cold atoms in an optical cavity, offers a test case for experimentally observing the emergent geometry of a quantum many-body system.
\end{abstract}

\maketitle

The fast scrambling conjecture---inspired by studies of the black-hole information problem---predicts a lower bound on the time for information to spread from one to all degrees of freedom of an $N$-body quantum system, scaling as $t_*\propto \log(N)$ \cite{sekino2008,hayden2007black}.  Fast scrambling is conceptually important as a putative signature of the quantum physics of black holes \cite{maldacena2015bound}.  It is also of practical importance because a fast scrambler is an efficient quantum encoder, capable of quickly entangling quantum information across many physical qubits \cite{Preskill:2016htv}.  While information scrambling has been probed in several pioneering experiments \cite{gaerttner2017measuring,li2017measuring,wei2018exploring,landsman2019verified}, observing \textit{fast} scrambling remains an outstanding challenge.

\begin{figure}[tb]
\includegraphics[width=\columnwidth]{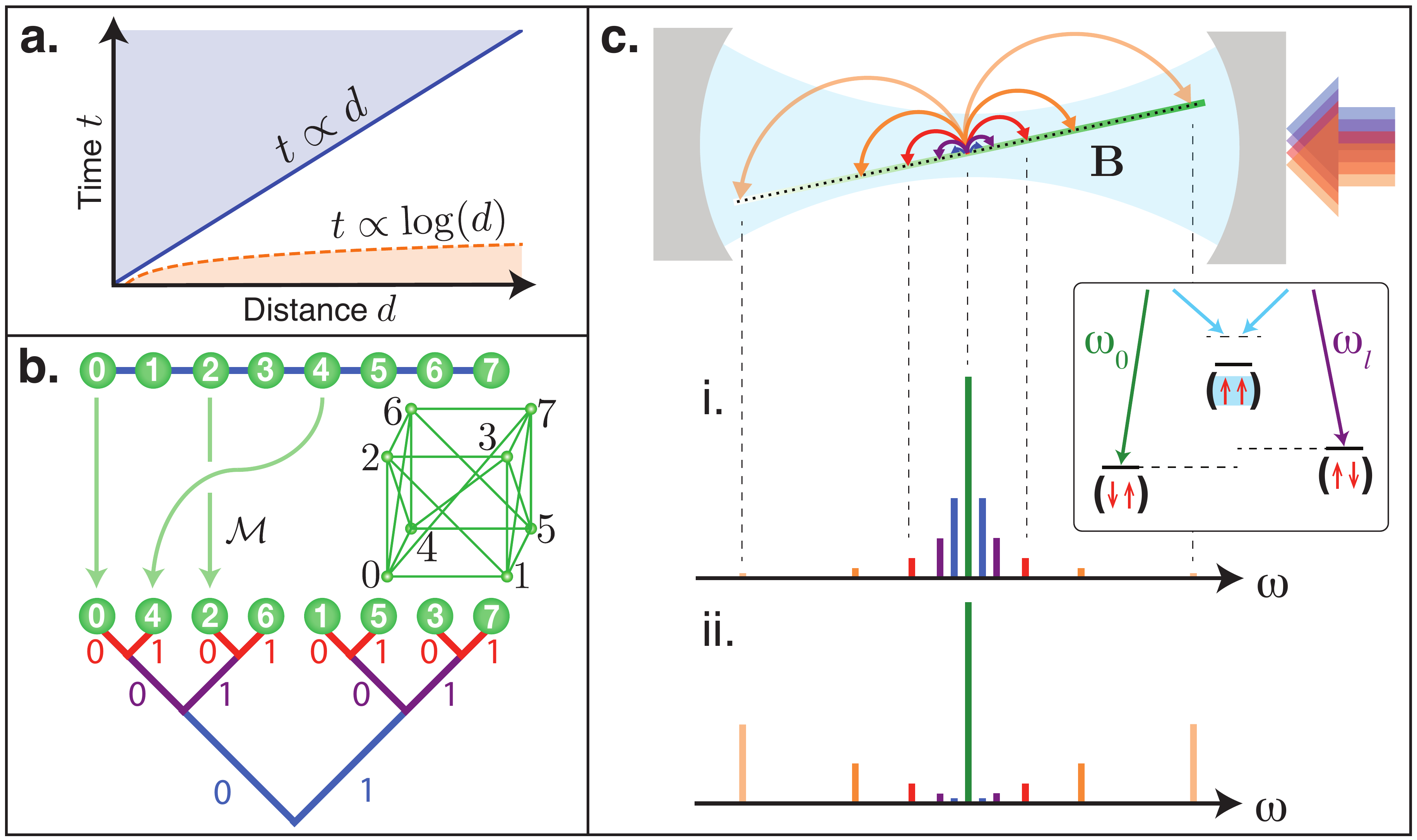}
\caption{\textbf{Nonlocal spin models with cold atoms.} (a) Linear light cone \vs fast scrambling.  (b) Spins (green) coupled locally along a chain of length $N=8$ or non-locally according to a tree of depth $n=\log_2(N)$.  Green cube shows full pattern of couplings specified in Eq.~\eqref{eq:Js}.  (c)  Scheme for controlling the graph of interactions mediated by a cavity via the spectrum of a drive field and the gradient of a magnetic field $\vec{B}$.  Depending whether short-range (blue) or long-range (orange) couplings dominate, the graph resembles either the linear chain (i) or the tree (ii).}\label{fig:overview}
\end{figure}

In typical quantum systems found in nature, fast scrambling is precluded by the locality of interactions.  In the presence of only short-range interactions, information propagation is bounded by a linear ``light cone'' \cite{lieb1972finite,cheneau2012light,richerme2014non,bravyi2006lieb} (Fig.~\ref{fig:overview}a), such that the scrambling time is at least $t_*\propto N$. Even with long-range power-law interactions \cite{hastings2006spectral,foss2015nearly,else2018improved}, the time for quantum correlations to propagate scales polynomially with distance \cite{foss2015nearly,else2018improved} and not logarithmically.  Potential work-arounds include engineering interactions among spatially overlapped modes \cite{danshita2017creating,chen2018quantum,chew2017approx} or using light to mediate effectively non-local interactions \cite{swingle2016measuring,marino2019cavity}.  Proposals to date have focused on emulating solvable models for fast scrambling featuring random all-to-all couplings \cite{danshita2017creating,chen2018quantum,chew2017approx,marino2019cavity}.

In this Letter, we propose an experimentally realizable quantum spin model featuring a sparse, deterministic graph of non-local interactions that enables fast scrambling.  A modest generalization of the simplest model provides a smooth interpolation between two radically different notions of geometry: linear or treelike, as shown in Fig.~\ref{fig:overview}b.  Our simplest model is, in a certain sense, exactly between these two limits.  Instead of exhibiting extended spatial geometry, it is fully characterized by a coupling graph whose diameter grows logarithmically with the number of sites, resulting in exponentially fast spreading of perturbations that decreases to power-law spreading as we tune away from the original model.    Thus, loosely speaking, we see the exponential spread of perturbations as a form of criticality encountered between two incompatible geometric notions of locality (Fig.~\ref{fig:overview}b).

In studying this transition in geometry, we are also motivated by holography, in particular the recently proposed $p$-adic version \cite{Gubser:2016guj,Heydeman:2016ldy} of the anti-de Sitter / Conformal Field Theory correspondence (AdS/CFT).  Here, the role of the gravitational bulk is played by the Bruhat-Tits tree, which is an infinite regular tree with $p+1$ edges leading into each vertex, similar to Fig.~\ref{fig:overview}b for $p=2$.  Past works have suggested that diffusion on the Bruhat-Tits tree can be used to understand aspects of fast scrambling near black hole horizons \cite{Barbon:2012zv,Barbon:2013goa}.

The models we consider here are inspired by an elegant proposal \cite{hung2016quantum} for engineering translation-invariant long-range interactions by encoding spins in cold atoms coupled to light in a waveguide or cavity \cite{gopalakrishnan2009emergent,leroux2010implementation,strack2011dicke,hosten2016quantum,welte2018photon,davis2019photon,kroeze2018spinor,norcia2018cavity,landini2018formation,braverman2019near,borjans2019longrange,mivehvar2019cavity}. This scheme can generate Hamiltonians of the form
\begin{equation}\label{eq:Hflipflop}
H = \frac{1}{2S}\sum_{i,j}J(i-j) S^+_i S^-_j,
\end{equation}
where $i,j = 1, \ldots, N$, and $\vec{S}_i$ is a spin-$S$ operator representing either an individual atom or the collective spin of a small atomic ensemble at site $i$.  We consider a sparse graph of couplings between sites separated by powers of two,
 \begin{equation}\label{eq:Js}
  J(i-j) = \begin{cases} 
     J_s 2^{\ell s} & \text{when}\,|i-j| = 2^\ell,\,\,\ell=0,1,2,3\dots \\
      0 & \text{otherwise},
   \end{cases}
 \end{equation}
where by tuning the exponent $s$ from $-\infty$ to $+\infty$ we interpolate between the linear and treelike limits.  The simplest choice $s=0$, where all non-zero couplings are equal, permits a perturbation to spread exponentially because the number of pairwise interactions required to get from site $i$ to site $j$ is bounded above by the number of binary digits in $i$ and $j$ that differ.

\begin{figure}[tbh]
\includegraphics[width=\columnwidth]{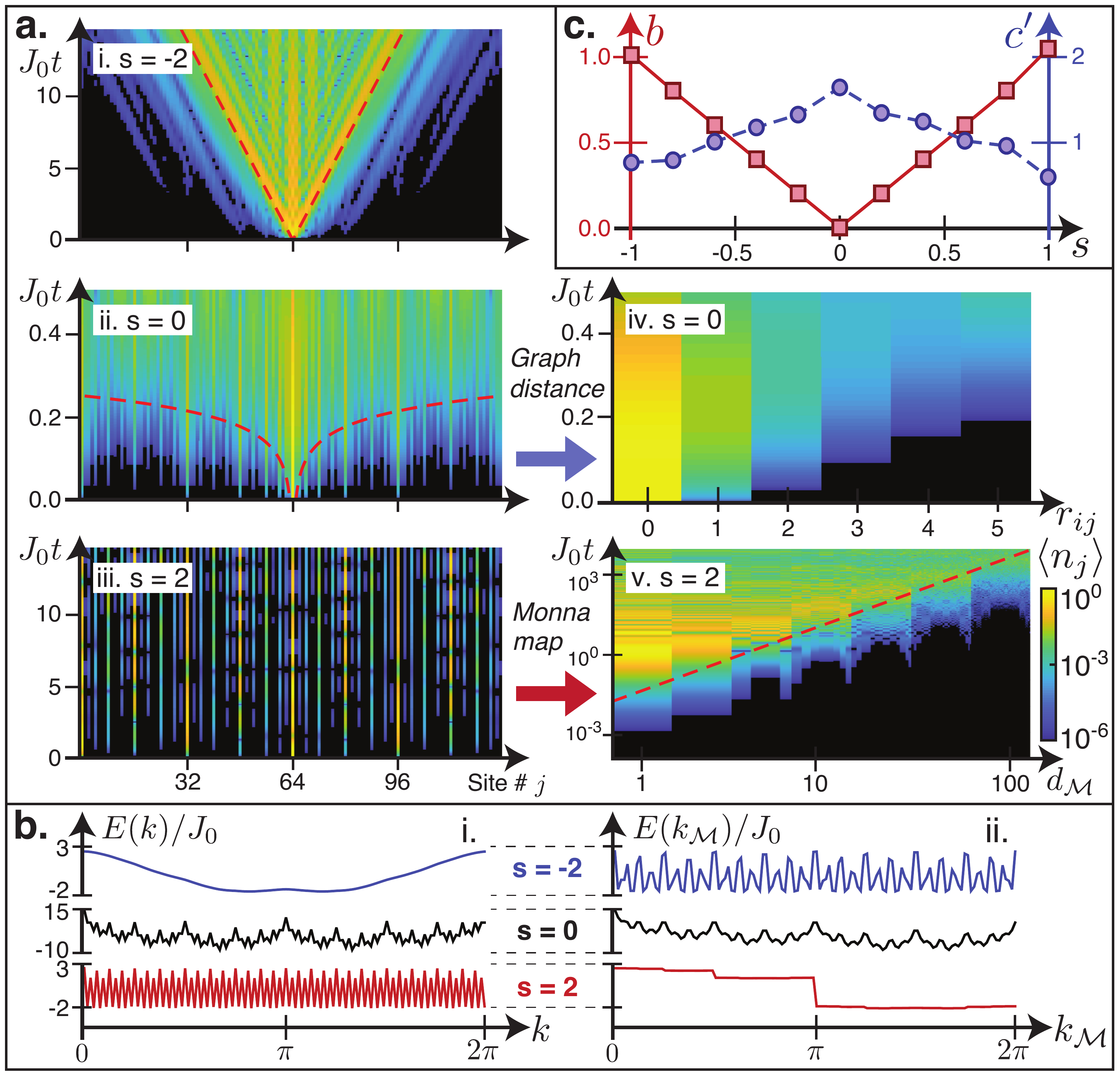}
\caption{\textbf{Dynamics in the single-magnon sector.} (a) Occupation $\avg{n_j(t)}$ for $N = 128$ sites initialized with a single excitation on site $N/2$ with (i) $s = -2$, (ii) $s = 0$, or (iii) $s=2$; slowest-growing occupations are bounded from above by dashed red lines. (iv) Occupation $\avg{n_j(t)}$ at $s = 0$ for $N=1024$, \vs graph distance $r_{ij}$ from initial site $i$. (v) Reemergence of light cone at $s=2$ after rearranging $N = 128$ sites according to Monna map $\mathcal{M}$. (b) Magnon dispersion relation for $N=128$ sites arranged according to either (i) physical location or (ii) Monna-mapped order for $s=-2$ (blue), $s=0$ (black), and $s=2$ (red). (c) Breakdown of polynomial light cone: polynomial exponent $b$ (red) and logarithmic exponent $c'$ (blue) versus $s$.
}\label{fig:single_magnon}
\end{figure}

The couplings in Eq.~\eqref{eq:Js} can be generated by a pair of Raman processes wherein one atom flips its spin by virtually scattering a photon from a control field into a cavity, and a second atom rescatters this photon such that its spin flops \cite{davis2019photon}.  In a magnetic field gradient \cite{hung2016quantum}, the energy cost of a flip-flop is proportional to the distance $d = i - j$ between spins.  Thus, inducing resonant flip-flops at distance $|d| = 2^\ell$ requires modulating the control field at a frequency $\omega_\ell \propto d$.  More generally, weakly modulating the control field at multiple frequencies $\omega_\ell$ for $\ell = 0, 1, 2, 3, \dots$ produces a set of sidebands whose amplitudes dictate the hopping amplitudes $J(d)$ (Fig.~\ref{fig:overview}).

Below, we take $N$ a power of 2 and assume periodic boundary conditions unless otherwise stated, letting $|x|$ denote the minimum value of $\sqrt{(x+qN)^2}$ over all integers $q$.  We normalize the couplings in Eq.~\eqref{eq:Js} such that the largest is always a constant $J_0$, letting $J_s = J_0$ for $s\leq 0$ and $J_s = J_0(N/2)^{-s}$ for $s>0$.

The key features of our spin model are evident already in the spreading of a single initially localized spin excitation (Fig.~\ref{fig:single_magnon}a), which is governed by the magnon dispersion relation (Fig.~\ref{fig:single_magnon}b)
 \begin{equation}\label{eq:disp}
  E(k) = 2 J_s \sum_{\ell=0}^{\log_2(N/2)} 2^{\ell s} \cos(2^\ell k) \,,
 \end{equation}
where $k \in [0,2\pi)$ is the wavenumber. Intriguingly, as $N$ increases, the dispersion relation shows fractal behavior converging to a Weierstrass function for $0 < s < 1$ \footnote{Related fractal behavior was recently found in a simplified statistical mechanical model with sparse couplings \cite{Gubser:2018bpe} of essentially the same form as in Eq.~\eqref{eq:Js}.}.   For arbitrary $s$, Eq.~\eqref{eq:disp} allows us to analytically compute the single-magnon dynamics as shown in Fig.~\ref{fig:single_magnon}a, where we introduce a single excitation on site $i = N/2$ and plot the mean occupation $\avg{n_{j}(t)}$ as a function of site number $j$ and time $t$.

The single-particle dynamics reveals the geometry of the interaction graph and its dependence on the exponent $s$.  When $s$ is large and negative, the excitation spreads ballistically, as expected for a nearest-neighbor spin chain, producing the linear light cone in Fig.~\ref{fig:single_magnon}a(i).  By contrast, for $s>0$, where interactions \textit{grow} with physical distance, the excitation jumps discontinuously between distant sites (Fig.~\ref{fig:single_magnon}a(iii)). Rather than interpreting the apparent absence of a light cone for $s>0$ as the absence of locality, we argue that a new version of locality emerges based on the $2$-adic norm $|x|_2 = 2^{-v(x)}$, where $2^{v(x)}$ is the largest power of 2 that divides $x$.  The distance $|i-j|_2$ between sites $i$ and $j$ is called ultrametric because the distance of the sum of two steps is never greater than the larger of the two steps' distance; by contrast the usual distance $|i-j|$ is called Archimedean because many small steps can be combined into a large jump.

We can understand the $2$-adic norm as a tree-like measure of distance because $|i-j|_2 = 2^{d_\text{tree}(i,j)/2}/N$, where $d_\text{tree}(i,j)$ is the number of edges between sites $i$ and $j$ along the regular tree in Fig.~\ref{fig:overview}b \footnote{The tree appears to break translation invariance, but in fact it does not: $d_\text{tree}(i+1,j+1) = d_\text{tree}(i,j)$.}.  The leaves are numbered in order of increasing ${\cal M}(i)$, where the discrete Monna map ${\cal M}$ reverses the bit order in the site number.  For example, for $N=8$ sites, ${\cal M}(1) = 4$ because in binary, ${\cal M}(001_2) = 100_2$.  Noting that $N k / 2 \pi$ is an integer, we may likewise define a Monna-mapped wavenumber $k_{\mathcal{M}}$ by
 \begin{equation}\label{eq:MonnaK}
  N \frac{ k_{\mathcal{M}}}{2\pi} = {\cal M}\left( N \frac{k}{2\pi}\right) \,.
 \end{equation}
For large positive $s$, we rearrange the spins according to the Monna map and find that a light cone reappears (Fig.~\ref{fig:single_magnon}a(v)) and the dispersion relation is smoothed out (Fig.~\ref{fig:single_magnon}b), corroborating the transformation to the treelike geometry defined by the $2$-adic norm.  

The radical difference between Archimedean and ultrametric geometry raises the question of what happens near $s=0$, where short- and long-range couplings are equally strong.  Here, it is useful to think of the sites as arranged on a hypercube \footnote{We thank P.~Hayden for discussions on this point.}, where each site number in binary specifies a corner.  The coupling graph consists of the edges of the hypercube plus some diagonals, as shown in Fig.~\ref{fig:overview}b for $N=8$ sites. For any $N$, the \textit{graph distance} $r_{ij}$ counts the minimum number of edges required to connect sites $i$ and $j$, and is at most $\lceil\frac{1}{2} \log_2(N) \rceil$, where $\lceil x \rceil$ denotes the smallest integer greater than or equal to $x$.

We expect the logarithmic diameter of the interaction graph at $s=0$ to enable a localized perturbation to spread over the entire system in a logarithmic time $t \propto \log(N)$. As a first test, in Fig.~\ref{fig:single_magnon}a we examine single-magnon transport as a function of both physical distance $d = |i-j|$ and graph distance $r_{ij}$. We observe spin transport on a timescale that is roughly linear in graph distance (Fig.~\ref{fig:single_magnon}a(iv)), and thus at most logarithmic in physical distance $d$ and system size $N$.

To permit this logarithmic timescale, the polynomial light cone in Fig.~\ref{fig:single_magnon}a(i) must break down as $s$ approaches 0. To examine this breakdown, we evaluate the time $t_{\epsilon}$ for the magnon occupation at a distance $d$ from the initial site to reach a threshold value $\avg{n_{i+d}} = \epsilon = 1/N^2$. For arbitrary $s$, we may always bound $t_{\epsilon}$ from below by a polynomial of the form $a \left[ d_{(\mathcal{M})} \right]^b \leq t_{\epsilon}$ that depends on physical distance $d$ for $s \leq 0$ or Monna distance $d_\mathcal{M} \equiv \mathcal{M}(d)$ for $s>0$, with non-negative constants $a,b$. We determine the dependence of the exponent $b$ on $s$ from a  fit to the points of fastest growth \cite{SM} in Fig.~\ref{fig:single_magnon}a, finding a direct proportionality $b\propto \abs{s}$ (red points in Fig.~\ref{fig:single_magnon}c). The vanishing exponent $b$ at $s=0$ signifies the breakdown of the polynomial light cone.
\nocite{Garcia-Ripoll2006}
\nocite{Paeckel2019}

Furthermore, we verify that even the \textit{slowest}-growing occupations take only a logarithmic time to reach the threshold $\epsilon$. For arbitrary $s$, the time $t_{\epsilon}$ is bounded above by $t_{\epsilon} \leq a' \left[ d_{(\mathcal{M})} \right]^{b'} \left[\log d_{(\mathcal{M})} \right]^{c'}$, with non-negative constants $a',b',c'$ (dashed red lines in Fig.~\ref{fig:single_magnon}a) \cite{SM}. Here, as $s \rightarrow 0$ the polynomial exponent $b'$ again disappears, but the logarithmic exponent $c'$ remains (blue points in Fig.~\ref{fig:single_magnon}c). For both the fastest- and slowest-growing occupations we observe an approximate symmetry about $s=0$ that suggests a duality between the Archimedean and non-Archimedean regimes \cite{Gubser:2018bpe}.

\begin{figure}[tb]
\includegraphics[width=\columnwidth]{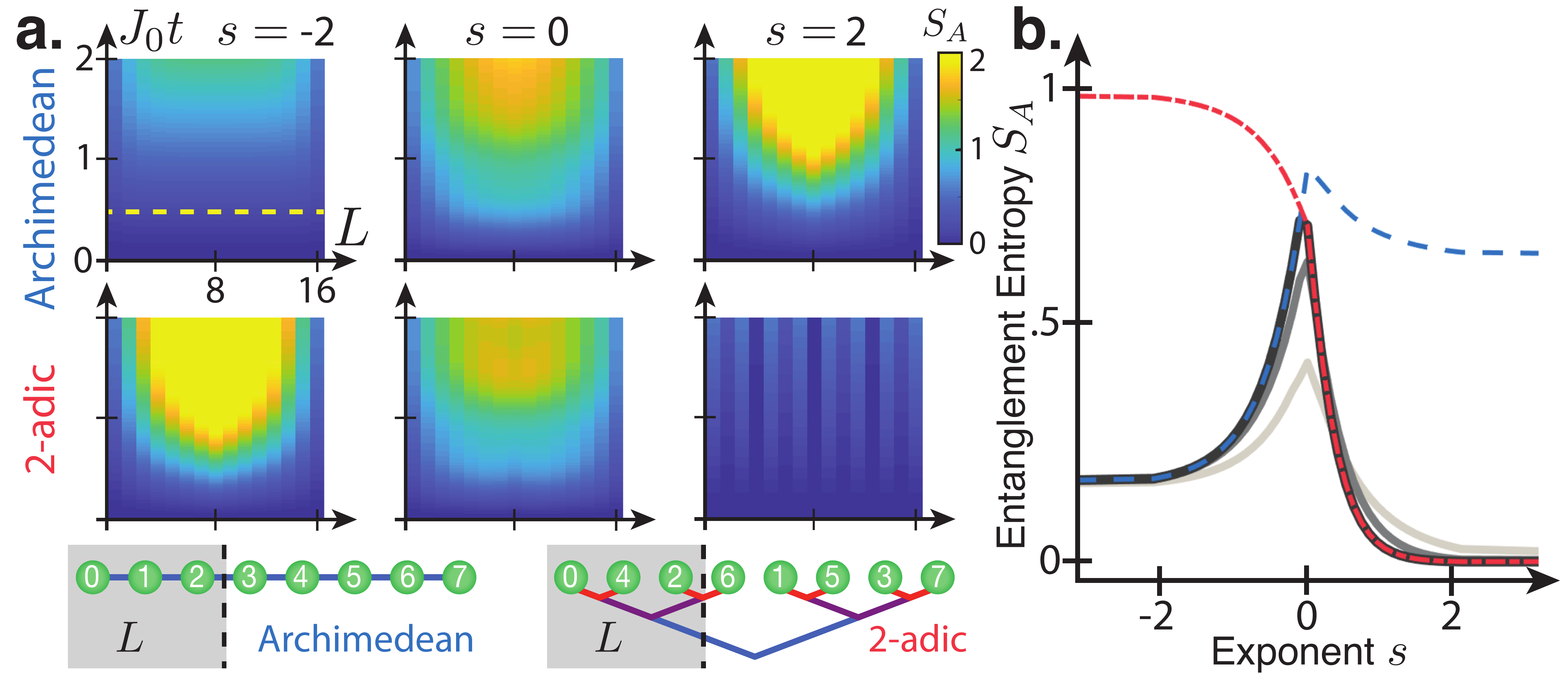}
\caption{\textbf{Growth of entanglement after a quench,} for an initial product state of $N=16$ sites with $S=1/2$ oriented along $\uvec{x}$.  (a) Entanglement entropy $S_A$ \vs partition size $L$ and time $t$, for Archimedean (top) or $2$-adic (bottom) partitions.  (b) Minimum entanglement \vs $s$ over {\it all} possible partitions of size $L=1,2,4,8$ (light to dark solid curves), compared with entanglement for Archimedean (blue dashed) and 2-adic (red dot-dashed) partitions.}
\label{fig:entanglement}
\end{figure}

While the spin dynamics provides strong evidence for the underlying geometry of the model as a function of $s$, the structure of entanglement provides an even sharper proxy for locality.  We analyze the growth of entanglement for a system initialized in a product state of spins polarized along $\uvec{x}$.  Figure~\ref{fig:entanglement}a shows the entanglement entropy $S_A$ of a subsystem $A$ as a function of its size and time, with the spins partitioned either according to their physical position (top) or to their Monna-mapped ordering (bottom).  In either of the two limits where $\abs{s}$ is large, there is a natural way of partitioning the system such that entanglement remains low, as summarized in Fig.~\ref{fig:entanglement}b: for $s<0$, the cut must be between nearest neighbors in the linear chain (blue curve), whereas for $s>0$ the cut must be between branches of the tree (red curve).

Near $s=0$, however, entanglement is high no matter how we cut the system.  We verify this by plotting the minimum entanglement entropy over {\it all} possible bipartitions, without regard to locality.  The resulting gray curves in Fig.~\ref{fig:entanglement}b, for different partition sizes $L$, show a sharp peak in entanglement at $s=0$.
Thus, at the crossover between the Archimedean and $2$-adic geometries, there is no good notion of locality, and all spins are strongly coupled to one another.

Our non-local spin models generically exhibit quantum chaos.  One indicator is that energy level spacings in the $s=0$ model at half filling ($N/2$ magnons for $S = 1/2$) exhibit random-matrix statistics \cite{SM,weinberg2017quspin,wimberger2014nonlinear}.  But is the highly connected model at $s=0$ a fast scrambler?  To probe this question, we consider the out-of-time-order correlation function (OTOC) \cite{larkin1969quasi,shenker2014black,swingle2016measuring,swingle2018resilience,zhang2019information}
\begin{equation}
    \label{eq:OTOC}
    C(i,j;t) = \avg{|[S^z_i(0),S^z_j(t)]|^2} / S^2 \,,
\end{equation}
where $S^z_{i,j}$ are local spin operators at sites $i,j$ and $S^z_j(t) = e^{i H t}S^z_j e^{-i H t}$. In a typical fast scrambler, $C$ approaches its saturation value on a timescale $t_* \propto \log(N)/\lambda$, where $\lambda>0$ is a Lyapunov exponent quantifying the system's exponential sensitivity to perturbations.

In systems with local interactions, fast scrambling is precluded by the Lieb-Robinson bound \cite{lieb1972finite,bentsen2019fast}, which restricts $C \lesssim e^{-(d-vt)}$ to exponentially small values at distances $d = \abs{i-j}$ outside a light cone with Lieb-Robinson velocity $v$, thereby preventing saturation of OTOCs until a time $t_* \propto N$.  By contrast, known models for fast scrambling feature random all-to-all couplings \cite{lashkari2011towards,maldacena2016remarks} and have no sense of spatial locality. Our model at $s=0$ offers an alternative route to fast scrambling: despite its effective light cone \vs graph distance $r_{ij}$ (Fig.~\ref{fig:single_magnon}a(iv)), the early-time growth of OTOCs is permitted to reach values $C \sim e^{-(\dgraph-v t)}\sim 1/N^\alpha$ \cite{bentsen2019fast} due to the logarithmic graph diameter $r_\mathrm{max} \approx \frac{1}{2} \log_2(N)$, where $\alpha$ is a constant of order unity and $v \propto J_0 \log(N)$ because each spin has $\log(N)$ couplings. Subsequent Lyapunov growth $C \sim e^{\lambda t} / N^{\alpha}$ therefore allows OTOCs to reach saturation in a time $t_* \sim \alpha \log (N) / \lambda$.


A Lyapunov regime, however, is not guaranteed or even expected in a finite-size system with small local Hilbert-space dimension, e.g., at spin $S=1/2$ \cite{khemani2018velocity}.  We thus first analyze the $s=0$ model semiclassically in the limit where each site contains a spin $S \gg 1$, which is natural to implement experimentally by letting each $\vec{S}_i$ represent the collective spin of an ensemble. We consider the averaged sensitivity
\begin{equation}\label{eq:sensitivity}
\C(i,j;t) = \frac{1}{S^2} \left\langle\left(\frac{d S^z_j(t)}{d \phi_i}\right)^2\right\rangle \,,
\end{equation}
for a small initial rotation $\phi_i$ of spin $i$ about the $z$-axis, whose correspondence to the OTOC can be observed by replacing the commutator in Eq.~\eqref{eq:OTOC} with a Poisson bracket \cite{polkovnikov2010phase,Cotler2018}.  We calculate the average in Eq.~\eqref{eq:sensitivity} for an ensemble in which each spin has a random initial orientation in the $xy$-plane, the classical limit of an infinite-temperature state at half filling.

\begin{figure}[tb]
\includegraphics[width=\columnwidth]{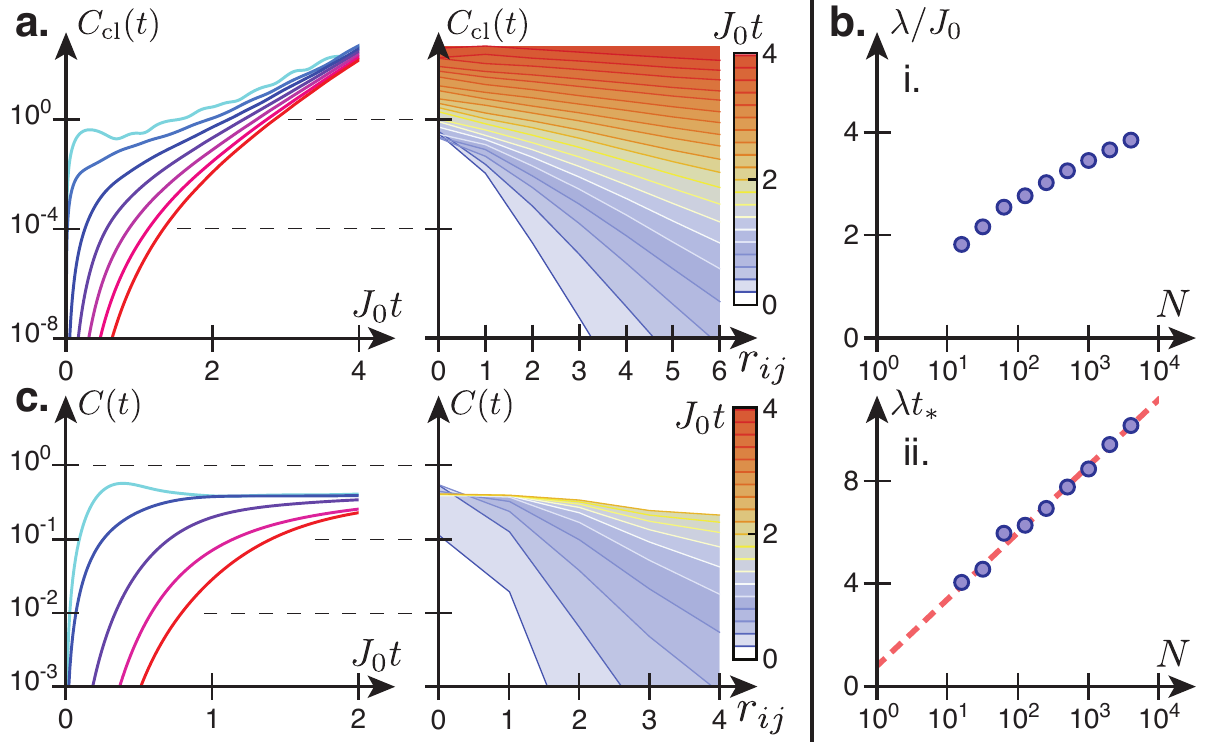}
\caption{
\textbf{Chaos and fast scrambling at $\mathbf{s=0}$.}
(a) Average semiclassical sensitivity $\C(t)$ for $N = 4096$. (i) $\C$ \vs time for graph distances $\dgraph = 0, 1, \dots 6$ (blue to red). (ii)  Fixed-time contours of $\C$ \vs $\dgraph$, showing exponential decay of $\C$ \vs $\dgraph$ at early times. Contours are for $0.2 \leq J_0 t \leq 4$ in increments of $0.2$. (b) Lyapunov exponent $\lambda / J_0$ (i) and scrambling time $\lambda t_* $ (ii) \vs $N$ (blue circles) with fit $\lambda t_* = \alpha \log(N) + \beta$ (red dashed). (c) MPS calculations for $N=64$, $S=1/2$, and open boundary conditions.}
\label{fig:TWA}
\end{figure}

The growth in sensitivity $\C(t)$ generically exhibits two distinct regimes \cite{marino2019cavity}, visible in Fig.~\ref{fig:TWA}.  The first is a rapid power-law growth $\C(t) \propto (J_0 t)^{2r_{ij}}$ for $J_0 t \lesssim 1$.  A transition to exponential growth occurs for $J_0 t \gtrsim 1$.  Crucial to fast scrambling is that, by the time the exponential growth begins, the OTOCs have already reached values $C_\mathrm{cl} \sim 1/N^{\alpha}$. We expect the subsequent exponential growth $\C(t)\sim e^{\lambda t}/N^{\alpha}$ to yield a value $\C(t_*) \sim 1$ at time $t_* \sim \alpha \log(N)/\lambda$.  To verify this behavior, we first fit the exponential growth for a range of system sizes to obtain the dependence of the Lyapunov exponent $\lambda$ on $N$, and then evaluate the time $t_*$ to reach $\C(t_*)=1$ in terms of $\lambda$ (Fig.~\ref{fig:TWA}b).  Fitting the dependence of this semiclassical scrambling time on $N$ yields $\lambda t_* = \alpha \log(N) + \beta$, with $\alpha=1.1(1)$.

Our semiclassical analysis demonstrates that the correlations developing at early times $J_0 t \lesssim 1$, \textit{before} the onset of exponential growth, are crucial for enabling fast scrambling: the weakest correlations should be only algebraically small ($C\propto N^{-\alpha}$) and not exponentially small in $N$.  We now apply this insight to investigate whether the quantum model at $S=1/2$ can be a fast scrambler.

As the essence of fast scrambling is efficient spreading of information across an exponentially large Hilbert space, the process is intrinsically difficult to study numerically.  Nevertheless, we calculate the early-time dynamics for $N=64$ sites at half filling using Matrix Product State (MPS) techniques \cite{Schollwoeck2011,Haegeman2011,Koffel2012,Haegeman2013,Haegeman2016,SM}, for a system at infinite temperature with open boundary conditions.  We see no Lyapunov regime, which is not surprising since there is no small parameter for $S=1/2$ at finite $N$ to prevent rapid saturation of the OTOC (Fig.~\ref{fig:TWA}c).  However, Fig.~\ref{fig:TWA}c(ii) indicates early-time correlations that fall off exponentially with graph distance $\dgraph$, consistent with correlations across all sites that are algebraically rather than exponentially small in $N$---the aforementioned necessary condition for fast scrambling.

Our results indicate that fast scrambling is accessible in sparsely coupled models without disorder, and might generalize to a wider range of coupling patterns in which the graph diameter grows logarithmically with system size.  Near-term cavity-QED experiments offer promise for observing the logarithmic timescale for spin transport at $s=0$, as well as the linear-to-treelike transition.  In the large-$S$ regime, where each site contains an atomic ensemble, experiments will benefit from a collective enhancement in the coherence of interactions.  Ultimately, implementations of the spin-1/2 model---for sufficiently strong atom-light coupling \cite{colombe2007strong,SM}---could test for fast scrambling \cite{SM,swingle2016measuring,swingle2018resilience,zhang2019information} at large $N \sim 10^3$ in the quantum regime.

Complementarily, future theoretical work may investigate whether a Lyapunov regime can emerge in the spin-1/2 quantum model through a coarse-graining procedure that exploits the self-similarity of the coupling pattern.  Measures of entanglement throughout the linear-to-treelike transition also merit further study, and may enable a more explicit connection to holography via entanglement wedges or tensor networks \cite{swingle2012constructing,pastawski2015holographic,bhattacharyya2018tensor,Heydeman:2016ldy,Heydeman:2018qty,Vidal2008,Shi2006}. Future work may also explore prospects for harnessing the rapid and deterministic generation of entanglement for quantum information processing.



\begin{acknowledgments}
\acknowledgments{This work was supported by the DOE Office of Science, Office of High Energy Physics.  MS-S acknowledges support from the Research Corporation Cottrell Scholar Program and the NSF.  SSG acknowledges support from the Simons Foundation. AJD, AB, and TH acknowledge support from the EPSRC Programme Grant DesOEQ (EP/P009565/1), and by the EOARD via AFOSR grant number FA9550-18-1-0064. We additionally acknowledge support from the NSF GRFP (EJD and GB) and Hertz Foundation (EJD), as well as support for international exchanges from SU2P. We thank Xiangyu Cao, Sean Hartnoll, Patrick Hayden, and Steve Shenker for helpful discussions.}
\end{acknowledgments}


\bibliography{scramble}

\clearpage
\onecolumngrid

\renewcommand{\thefigure}{S\arabic{figure}}
\renewcommand{\theequation}{S\arabic{equation}}

\setcounter{equation}{0}
\setcounter{figure}{0}

\section{Treelike Interactions and Fast Scrambling with Cold Atoms:\\Supplemental Material}

In this supplement, we elaborate on the proposed experimental implementation and on numerical and analytical techniques used to obtain the results in the main text. In Section I we discuss details of the experimental implementation including the effects of dissipation. In Section II we explain the methods used to find upper and lower bounds on magnon occupation growth times $t_{\epsilon}$, present evidence for chaotic level statistics, discuss the short-time expansion of OTOCs, and elaborate on the semiclassical and Matrix-Product State numerical methods used in the main text.

\section{I. Experimental Implementation}

\subsection{A. Engineering the Couplings}

We propose an experimental implementation in a one-dimensional array of atoms uniformly coupled to the mode of an optical cavity, with vacuum Rabi frequency $2g$.  At each of $N$ sites are $n$ atoms that encode a spin of length $S = n F$, either in the Zeeman states of an individual spin-$F$ atom ($n=1$) or in the collective magnetization of a localized ensemble.  A magnetic field gradient produces a Zeeman splitting $\omega_Z + j\omega$ that depends linearly on the site number $j\in \{1,\dots,N\}$.  Spin-exchange interactions are introduced by an optical control field at large detuning $\Delta$ from atomic excited states, as in Ref. \cite{davis2019photon}.  This control field, with center frequency $\omega_0$, drives virtual Raman processes wherein an atom absorbs a photon and emits it into the cavity mode at frequency $\omega_c$, whence it is rescattered by another atom.

\begin{figure}[hb]
\includegraphics[width=8cm]{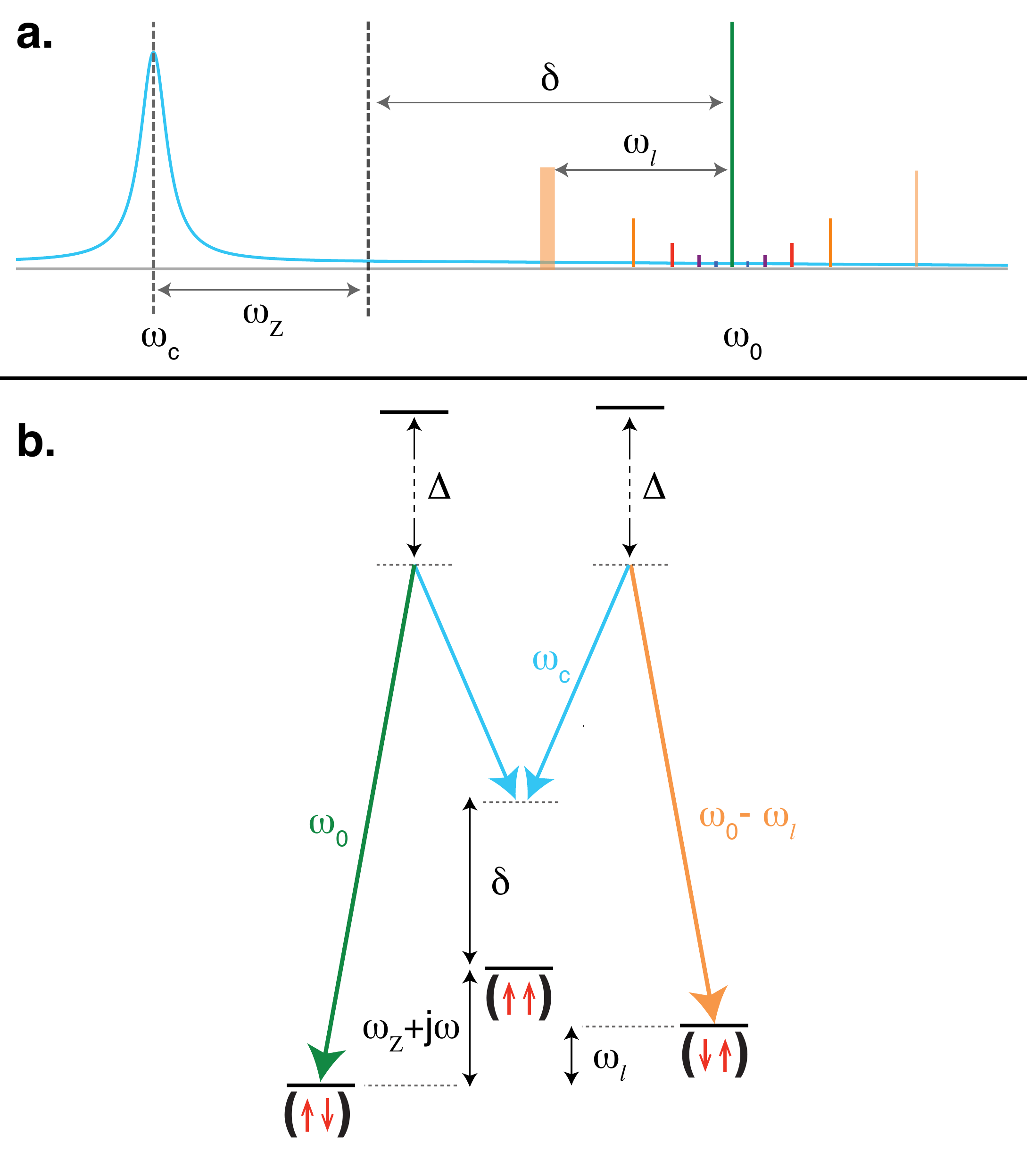}
\caption{\textbf{Experimental implementation.} (a) Spectrum of frequencies: cavity resonance, Raman resonance, and a set of drive fields for $s>0$ are sketched. The drive field corresponding to the particular four-photon process depicted in (b) is shown in bold.  (b) Photon-mediated spin-exchange in a magnetic field gradient. A photon is absorbed from the carrier at $\omega_0$ and exchanged with another atom via the cavity mode at frequency $\omega_c$, followed by emission into a sideband at $\omega_0 - \omega_\ell$.}\label{fig:expSchematic}
\end{figure}

To turn on interactions at each distance $d = 2^\ell$, the control field is amplitude-modulated at frequency $\omega_\ell$ and modulation index $\beta_\ell \ll 1$ to generate sidebands of Rabi frequencies $\Omega_\ell = \beta_\ell \Omega_0$, where $\Omega_0$ is the Rabi frequency of the carrier.  We assume that the detuning $\delta = \omega_0 - \omega_c - \omega_Z$ of the carrier from Raman resonance is large compared to all modulation frequencies and to the cavity linewidth $\kappa$, i.e., we assume $\delta > (\log_2 N) \omega$ and $\delta > \kappa$. The spin-exchange couplings are then given by
\begin{equation}
J(d=2^\ell) \propto \frac{n\Omega_\ell\Omega_0 g^2}{\Delta^2 \delta},
\end{equation}
where $n$ is the number of atoms per site.

\subsection{B. Effects of Dissipation}

To calculate how coherent the interactions can be, we would like to compare the couplings $J$ with the rates of two decay processes: collective decay via photon loss from the cavity mode and single-atom decay due to spontaneous emission.  To this end, we start by re-expressing the interaction strengths in terms of the free-space scattering rate due to the control field carrier,
\begin{equation}
\GscC = \frac{\Omega_0^2}{4\Delta^2}\Gamma,
\end{equation}
where $\Gamma$ is the atomic excited-state linewidth.  In terms of $\GscC$ and the single-atom cooperativity $\eta = 4g^2/(\kappa \Gamma)$, we have
\begin{equation}\label{eq:Jgsc}
J(d=2^\ell) \sim \beta_\ell n\eta \frac{ \GscC\kappa}{\delta},
\end{equation}
which shows that interactions can dominate over spontaneous emission for sufficiently large cooperativity $n\eta$. Below, in identifying requirements on the cooperativity, we will keep in mind that the total spontaneous emission rate is augmented by the power in the sidebands: $\Gamma_\mathrm{sc} = B\GscC$, where $B = 1+ 2\sum_\ell \abs{\beta_\ell}^2$.

A second dissipation mechanism is photon loss from the cavity. Cavity decay leads to collective dissipation described by a set of Lindblad operators
\begin{equation}
L_{k,\pm} = \sqrt{\gamma_{k,\pm}} S^\pm_k,
\end{equation}
where we have defined the spin-wave operators
\begin{equation}
\vec{S}_k = \frac{1}{\sqrt{N}}\sum_{j=1}^N e^{i k x_j} \vec{S}_j.
\end{equation}
The decay rates $\gamma_{k,\pm}$ are related to the dispersion relation $E_k$ in Eq.~\eqref{eq:disp} of the main text by
\begin{equation}
\gamma_{k,\pm} \sim \frac{\kappa}{\delta}E_k.
\end{equation}
For example, for our $s=0$ model, which has $\log_2(N/2)$ couplings of equal strength $J_\ell \equiv J$, the decay rates $\gamma_{k,\pm}$ are bounded above by
\begin{equation}
\gamma \lesssim \log_2(N/2)\frac{J\kappa}{\delta}.
\end{equation}

\subsection{C. Interaction-to-Decay Ratio}

We calculate an overall interaction-to-decay ratio at $s=0$ by comparing the couplings $J$ with the rates of collective decay $\gamma$ via the cavity and spontaneous emission $\Gsc$.  We thus obtain a ratio
\begin{equation}
\rho \equiv \frac{J}{\gamma + B\GscC} \sim \frac{\beta}{\frac{M\beta\kappa}{\delta} + \frac{B\delta}{n\eta\kappa}},
\end{equation}
where $\beta$ is the modulation index for each sideband, $M \equiv \log_2(N/2)$ is the number of modulation frequencies, and $B=1+ 2M\beta^2$.  At an optimal detuning $\delta/\kappa \sim \sqrt{n\eta M\beta/B}$, we obtain
\begin{equation}\label{eq:rho}
\rho = \frac{1}{2}\sqrt{\frac{n\eta\beta}{MB}} = \frac{1}{2}\sqrt{\frac{n\eta\beta/M}{1 + 2 M \beta^2}}.
\end{equation}
Eq.~\eqref{eq:rho} is maximized by choosing the modulation index such that total power in the sidebands equals the power in the carrier, $2M\beta^2 = 1$, which is consistent with the weak-modulation requirement $\beta\ll 1$ provided that $N$ is large.  The interaction-to-decay ratio is then
\begin{equation}\label{eq:rho2}
\rho = \frac{\sqrt{n\eta}/2}{(2M)^{3/4}}.
\end{equation}

\begin{figure}[h]
\includegraphics[width=0.5\columnwidth]{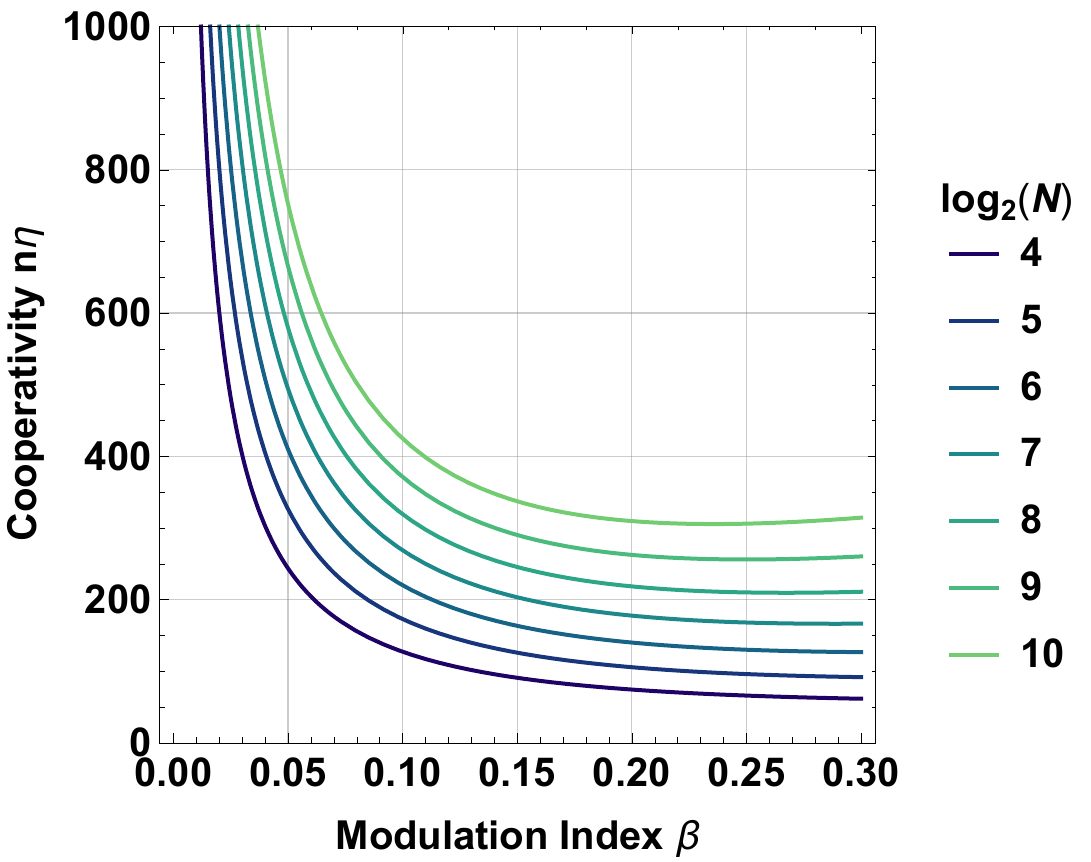}
\caption{\textbf{Cooperativity required to reach the scrambling time}, corresponding to an interaction-to-decay ratio $\rho = 1$.  The required cooperativity $n\eta$ per subensemble is plotted as a function of modulation index $\beta$ for system sizes ranging from $N=2^4$ to $N=2^{10}$ sites.}\label{fig:coop}
\end{figure}

Reaching the scrambling time $Jt \sim 1$ requires an interaction-to-decay ratio $\rho \gtrsim 1$.  By Eq.~\eqref{eq:rho2}, we thus require a large collective cooperativity of the subensemble at each site:
\begin{equation}\label{eq:neta}
\frac{n\eta}{4} \gtrsim [2\log_2(N/2)]^{3/2},
\end{equation}
where the weak scaling of the right-hand side with $N$ is convenient for scaling to large system sizes.  For example, Eq.~\eqref{eq:rho2} can be satisfied for a system of $N = 2^{10}$ sites with $n=300$ atoms per site and single-atom cooperativity $\eta \sim 1$ (Fig. \ref{fig:coop}).  A higher but demonstrated single-atom cooperativity $\eta \sim 150$ \cite{colombe2007strong} suffices to begin exploring fully quantum spin-1/2 models with a single atom per site.

Even when these conditions on the cooperativity are satisfied, dissipation may modify shape of the light cone, as pointed out in Ref. \cite{zhang2019information}. This effect can be mitigated by measuring `corrected' OTOCs \cite{zhang2019information,swingle2018resilience}, which divide out the effects of dissipation and recover the shape of the true light cone.

\subsection{D. Effects of Finite Modulation Depth}

Our derivation of the couplings $J_\ell$ in Sec. I.A. assumes the conceptually simple limit of small modulation depth $\beta_\ell \ll 1$.  In practice, for the system sizes most readily accessible in experiments, a finite modulation depth $\beta \sim 0.2$ is advantageous for maximizing the interaction-to-decay ratio (Fig. \ref{fig:coop}).   Operating at finite modulation depth adds weak additional couplings (a factor $\beta$ smaller than the intended couplings) at distances that are not powers of two.  While these couplings are unlikely to appreciably alter the fast scrambling dynamics or the geometry of the coupling graph, it is worth noting that the modulation waveform can be designed to produce only the desired couplings even at finite modulation depth.  In the most general case, for a control field $\Omega(t)$, the amplitude modulation waveform $|\Omega(t)|^2$ dictates the magnon dispersion relation $E(k) \propto |\Omega(k/\omega)|^2$.  Thus, while the field $\Omega(t) \propto 1 + 2\sum_\ell \beta_\ell \cos(2^\ell \omega t)$ produces precisely the desired coupling pattern in the limit $\beta_\ell \ll 1$, a more general class of waveforms $\Omega(t) \propto \sqrt{E(\omega t) + \mathrm{const.}}$ produce the same coupling pattern at finite modulation depth.

\section{II. Numerical and Analytical Techniques}

\subsection{A. Magnon Occupation and Information Spreading}

To characterize the spreading of local perturbations in the single-magnon sector, we introduce a single excitation at site $i$ and evaluate the time $t_{\epsilon}$ at which the magnon occupation at another site $j$ reaches a threshold value $\avg{n_{j}} = \epsilon = 1/N^2$. We note that the magnon occupation $\avg{n_j(t)}$, when the system is initialized with one excitation at site $i$, is equivalent to the OTOC
\begin{equation}
    \avg{n_j(t)} = \bra{0} c_i \adj{c}_j(t) c_j(t) \adj{c}_i \ket{0} = \bra{0} [c_i, \adj{c}_j(t)] [c_j(t), \adj{c}_i] \ket{0}
\end{equation}
since $c_j(t) \ket{0} = 0$. (Analogous relations hold for more general lowering operators $A^-,B^-$, so long as $B^- \ket{0} = 0$ and $\ket{0}$ is an eigenstate of $H$.) The occupation $\langle n_j(t) \rangle$ therefore encodes similar information about the spread of information in the system and is bounded by the Lieb-Robinson bound that controls the OTOC.


\subsection{B. Characterizing Fastest- and Slowest-Growing Magnon Occupations}

\begin{figure}[h]
\includegraphics[width=0.8\columnwidth]{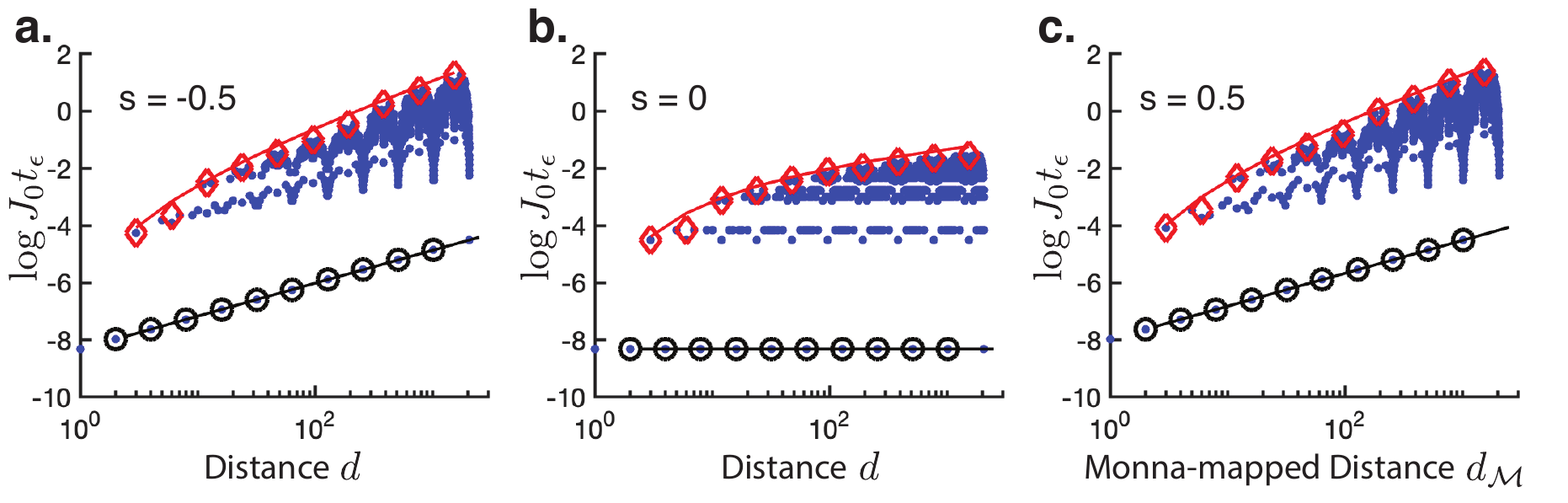}
\caption{\textbf{Upper and lower bounds on single-magnon occupation.} Time $t_{\epsilon}$ required for average magnon occupation $\avg{n_{i+d}(t)}$ to reach a fixed threshold $\epsilon$ vs physical distance $d$ or Monna-mapped distance $d_{\mathcal{M}}$ for (a) $s = -0.5$, (b) $s = 0$, and (c) $s = 0.5$. Fastest-growing occupations (black circles) are bounded from below by a power-law in $d_{(\mathcal{M})}$ (black lines), while slowest-growing occupations (red diamonds) are bounded from above by the function (\ref{eq:boundabove}) (red lines).}\label{fig:fastslowcorr}
\end{figure}

In Fig.~\ref{fig:fastslowcorr}, we plot the threshold time $t_{\epsilon}$ versus physical distance $d = |i-j|$ or Monna-mapped distance $d_{\mathcal{M}} = \mathcal{M}(d)$ from the original site for different values of $s$. While these plots show significant scatter, we observe that one can always bound the values $t_{\epsilon}$ from below by a polynomial light cone of the form
\begin{equation}
    \label{eq:boundbelow}
    a \left[ d_{(\mathcal{M})} \right]^b \leq t_{\epsilon}
\end{equation}
and from above by a function of the form 
\begin{equation}
    \label{eq:boundabove}
    t_{\epsilon} \leq a' \left[ d_{(\mathcal{M})} \right]^{b'} \left[\log d_{(\mathcal{M})} \right]^{c'}
\end{equation}
for some non-negative constants $a,b,a',b',c'$. To choose the lower-bound coefficients $a,b$ optimally, we first fit Eq.~\eqref{eq:boundbelow} to the fastest-growing occupations (black circles in Fig.~\ref{fig:fastslowcorr}), which always occur at distances $d_{(\mathcal{M})}$ that are powers of 2. The resulting lower bounds are shown in black in Fig.~\ref{fig:fastslowcorr}.

To find optimal upper-bound coefficients $a',b',c'$, we first isolate the slowest-growing occupations by splitting the scatter plot into bins $d \in [2^n,2^{n+1})$ for integers $0 \leq n \leq \log_2 N - 2$ and picking the largest value of $t_{\epsilon}$ within each bin (red diamonds in Fig.~\ref{fig:fastslowcorr}). We then use a constrained minimization search to find coefficients $a',b',c'$ that minimize the sum of the squares of the differences between the slowest data points in each bin and the upper-bound function of Eq.~\eqref{eq:boundabove}. In addition, to ensure that the resulting function bounds the slowest points from above, we place a constraint on the minimization search such that the optimal function must always lie \textit{above} the slowest datapoints. Finally, to ensure that the upper and lower bounds Eqs.~\eqref{eq:boundbelow} and \eqref{eq:boundabove} hold simultaneously for large $d_{(\mathcal{M})}$, we demand that $a' \geq a$ and $b' \geq b$. The resulting optimal upper-bound curves are plotted in red in Fig.~\ref{fig:fastslowcorr}.

\subsection{C. Energy Level Statistics}

\begin{figure}[h]
\includegraphics[width=0.5\columnwidth]{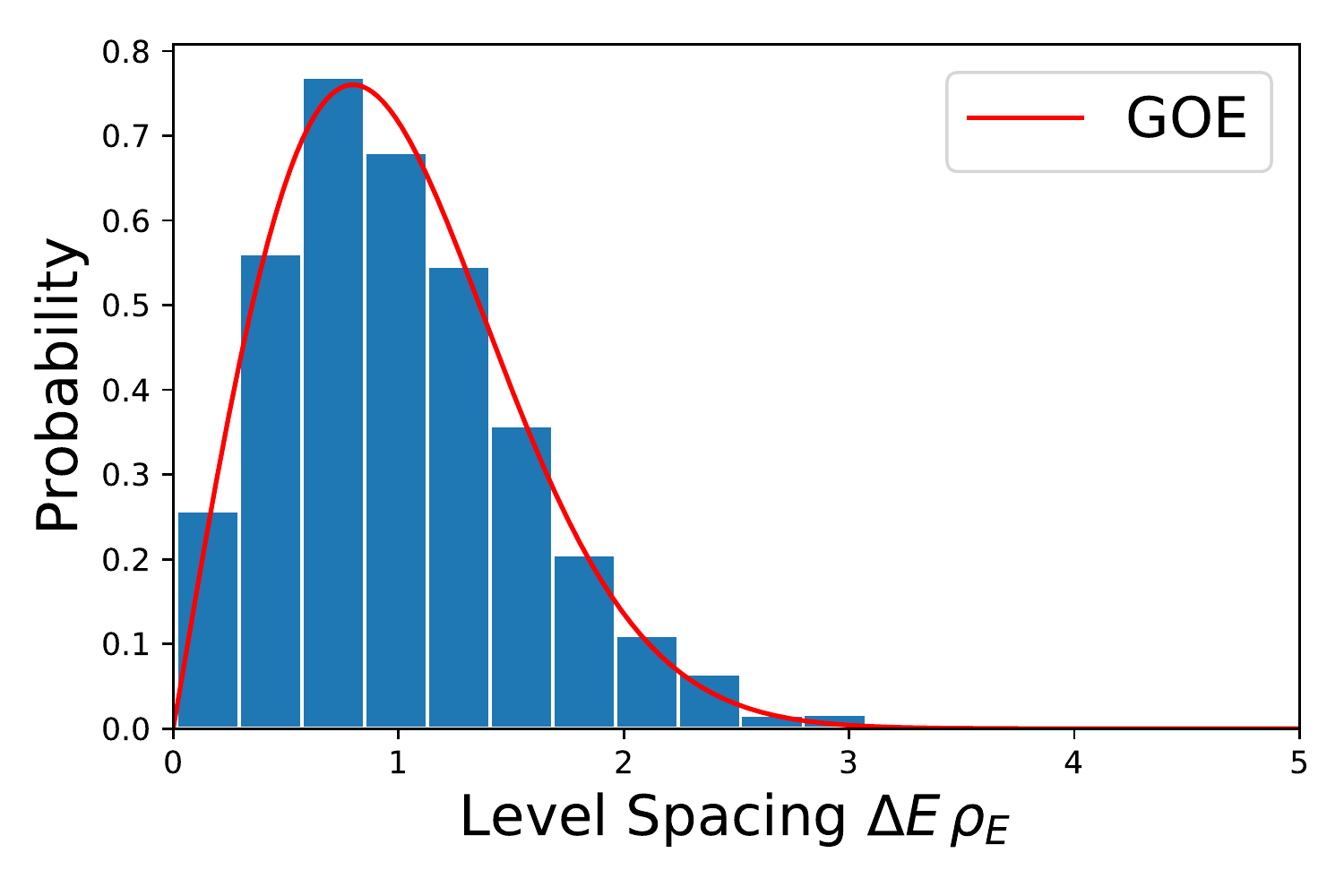}
\caption{\textbf{Chaotic level statistics.}  Distribution of spacings between neighboring energy levels for the $s=0$ model with $N=16$ sites at half filling ($N/2$ magnons).  The level spacings $\Delta E$ calculated for each momentum sector are normalized according to a Gaussian fit to the density of states $\rho_E$.  Red curve shows the Wigner-Dyson distribution for the Gaussian Orthogonal Ensemble, indicating that the interacting system exhibits random-matrix statistics.}\label{fig:DeltaE}
\end{figure}

To verify that the $s=0$ model is chaotic in the interacting quantum regime, we calculate the energy level statistics for a system of $N=16$ sites at half filling by exact diagonalization \cite{weinberg2017quspin}.  Figure \ref{fig:DeltaE} shows the distribution of spacings between neighboring energy levels, after unfolding the spectrum in each momentum sector according to the procedure in Ref. \cite{wimberger2014nonlinear}.  The distribution is consistent with the random-matrix statistics of the Gaussian Orthogonal Ensemble.

\subsection{D. Short-Time Expansion of OTOCs}

Here we show that OTOCs display power-law growth $C(t) \propto (J t)^{2 r_{ij}}$ at early times $J t \ll 1$ \cite{marino2019cavity}, where $J$ represents the typical energy scale of the Hamiltonian $H$. Consider the OTOC $C(i,j;t) = \langle [\Op_i(t), \Op_j(0)]^2 \rangle$, where $\Op_i$ is an arbitrary operator at site $i$. At early times, the growth of OTOCs can be understood via a short-time expansion:
\begin{equation}
    \Op_i(t) = \Op_i(0) - i t [\Op_i, H] - \frac{1}{2} t^2 [[\Op_i,H],H] + \cdots
    \label{eq:ShortTimeExpansion}
\end{equation}
If there is a direct coupling between sites $i$ and $j$, then at short times $t \ll J^{-1}$ the above expansion is simply dominated by the first two terms, leading to:
\begin{equation}
    \langle [\Op_i(t), \Op_j(0)]^2 \rangle \approx t^2 \langle [[\Op_i, H_{ij}], \Op_j]^2 \rangle
\end{equation}
where $H_{ij}$ is the term in the Hamiltonian coupling sites $i \neq j$. If there is no direct coupling between $i,j$, then this lowest-order quadratic term vanishes and we must consider higher-order terms in the expansion. More generally, if $r_{ij}$ is the minimum number of couplings $H_{nm}$ required to hop from site $i$ to site $j$, then all lowest-order terms $\sim t^{2n}$ for $n < r_{ij}$ vanish, and the early-time growth is dominated by
\begin{equation}
    C(i,j;t) \propto (J t)^{2 r_{ij}}
\end{equation}
where the constant of proportionality includes contributions from nested commutators $\langle [[[[\Op_i, H], H], \ldots, H], \Op_j]^2 \rangle$. For example, for the interaction graph at $s = 0$, we find that $C(0,1;t)$ and $C(0,2;t)$ grow like $(J_0 t)^2$ at early times since there is a direct coupling between sites $0,1$ and sites $0,2$. Similarly, $C(0,3;t)$ grows like $(J_0 t)^4$ because it requires two couplings to hop between sites 0 and 3, and $C(0,11;t)$ grows like $(J_0 t)^6$ because it requires three couplings to hop between sites 0 and 11.

\subsection{E. Semiclassical Numerics}

While exact diagonalization and Matrix Product State techniques can give us access to the full quantum dynamics of the Hamiltonian (1), these techniques limit us to relatively small system sizes. To examine information spreading in larger system sizes while retaining access to strong interactions and chaos, we must resort to a semiclassical treatment by considering the limit where the spin lengths $S \rightarrow \infty$ \cite{polkovnikov2010phase}. In this limit, each spin operator $\vec{S}_i$ can be understood as a classical angular momentum vector obeying the Poisson bracket algebra:
\begin{equation}
    \left\{ S_i^{\alpha}, S_j^{\beta} \right\} = \sum_{\gamma} S_i^{\gamma} \epsilon^{\alpha \beta \gamma} \delta_{ij}
\end{equation}
and whose equations of motion are given by
\begin{equation}
    \dot{S}_i^{\alpha} = \left\{S_i^{\alpha}, H \right\}\,.
\end{equation}
Defining unit-length spin vectors $\vec{x}_i = \vec{S}_i/S$ and taking the limit $S \rightarrow \infty$, we obtain classical equations of motion of the form:
\begin{equation}
    \dot{\vec{x}}_i = f(\vec{x}_i)
\end{equation}
where $f$ is a quadratic function of the spin vectors $\vec{x}_i$ that depends on the coupling graph $J(i-j)$ but is independent of $S$.

To mimic an infinite-temperature quantum state consisting of exactly $N/2$ magnons, we prepare the spins along random directions in the $x-y$ plane. (Preparing spins with additional out-of-plane fluctuations in the $z$-direction, or with uniformly random directions, has no significant effect on the dynamics.)
We then estimate the classical sensitivity $C_{\mathrm{cl}}(t)$ by perturbing a single spin at site $i$ by a small rotation $\phi_i$ about the $z$ axis, and measure the resulting change $\Delta \vec{x}_j(t)$ in spin $j$ at a later time $t$.

\subsection{F. Simulating OTOCs with Matrix Product Operators}

Here we describe how OTOCs are calculated using tensor network methods. We perform simulations with Matrix Product States and Operators (MPS/MPO) \cite{Schollwoeck2011}, which a regularly used to successfully simulate quantum many-body dynamics in one dimensional systems \cite{Garcia-Ripoll2006, Paeckel2019}. 
Although here we have only made use of MPS and MPO techniques, in future work it might be beneficial to try other generalized forms of tensor networks for treating these sparse models, such as the Tree Tensor Network \cite{Shi2006} or the Multi-scale Entanglement Renormalization Ansatz \cite{Vidal2008}. These may account more naturally for the geometry of long-range sparse interactions between spins.

In our calculations, we consider the symmetry sector of the Hilbert space of $N$ spins with a fixed number of magnons, $n$.
The initial state of the system $\rho_0$ is the infinite-temperature state within this symmetry sector, and may be written as $\rho_0 = P_n / {\mathcal Z}$, where the $P_n$ is the projector on states with $n$ magnons and ${\mathcal Z}$ is the partition function, which in the case of infinite temperature is equal to the number of permutations of $n$ magnons on $N$ sites.

We construct the projector $P_n$ as an MPO by writing it as a product of $n \times n$ matrices,
\begin{equation}
P_n = {\mathcal B}_n^{[1]} {\mathcal B}_n^{[2]} ... {\mathcal B}_n^{[N]},
\end{equation}
where
\begin{equation}
\begin{array}{l}
{\mathcal B}_n^{[1]}=
\left(\begin{array}{ccccc}
p_\downarrow & p_\uparrow & 0 & 0 & \ldots 
\end{array}\right),
\\
{\mathcal B}_n^{[1<i<N]}=
\left(\begin{array}{ccccc}
p_\downarrow & p_\uparrow & 0 & 0 & \ldots \\
0 & p_\downarrow & p_\uparrow & 0 &  \\
0 & 0 & p_\downarrow & p_\uparrow & \\
0 & 0 & 0 & p_\downarrow  & \\
\vdots & & & & \ddots
\end{array}\right), 
\\
{\mathcal B}_n^{[N]}=
\left(\begin{array}{c}
\vdots\\
0\\
0\\
p_\uparrow\\
p_\downarrow
\end{array}\right),
\end{array}
\end{equation}
are constructed out of operators $p_\downarrow=\ket{\downarrow} \bra{\downarrow}$ and $p_\uparrow=\ket{\uparrow} \bra{\uparrow}$ acting on local spins.

Having constructed this infinite temperature state, the next step is to time evolve the operator $S_j^z(t)$ from Eq.~\eqref{eq:OTOC} in the Heisenberg picture. In order to do this, we combine the local indices of the operator, thus vectorizing it so that it can be propagated in time using the same techniques we apply for MPS.
We find that using a TDVP \cite{Haegeman2011, Koffel2012, Haegeman2013, Haegeman2016} provides a good balance between speed and accuracy in this calculation (although for some purposes, using the Runge-Kutta method \cite{Garcia-Ripoll2006} could yield higher accuracy).
After the time evolution, we separate out the local indices of $S_j^z(t)$ and contract it with $S_i^z(0)$ and $\rho_0$ to obtain $C(i,j;t)$ in Eq.~\eqref{eq:OTOC}.

\end{document}